\documentclass[aps,prl,reprint,groupedaddress,superscriptaddress,showpacs,amsmath]{revtex4-1}
\usepackage{graphicx}
\usepackage{ulem}

\draft 

\begin{document}
\squeezetable
\title{Nucleation at the contact line observed on nano-textured surfaces} 



\author{C. W. Gurganus}
\affiliation{Department of Physics, Michigan Technological University, Houghton, Michigan 49931, USA}
\affiliation{Atmospheric Sciences Program, Michigan Technological University, Houghton, Michigan 49931, USA}

\author{J. C. Charnawskas}
\affiliation{Department of Physics, Michigan Technological University, Houghton, Michigan 49931, USA}

\author{A. B. Kostinski}
\affiliation{Department of Physics, Michigan Technological University, Houghton, Michigan 49931, USA}
\affiliation{Atmospheric Sciences Program, Michigan Technological University, Houghton, Michigan 49931, USA}

\author{R. A. Shaw}
\email[Correspondence to: ]{rashaw@mtu.edu}
\affiliation{Department of Physics, Michigan Technological University, Houghton, Michigan 49931, USA}
\affiliation{Atmospheric Sciences Program, Michigan Technological University, Houghton, Michigan 49931, USA}



\date{\today}

\begin{abstract}
\paragraph*{}
It has been conjectured that roughness plays a role in surface nucleation, the tendency for freezing to begin preferentially at the liquid-gas interface. Using high speed imaging, we sought evidence for freezing at the contact line on catalyst substrates with imposed characteristic length scales (texture). Length scales consistent with the critical nucleus size and with $\delta \sim \tau/\sigma$, where $\tau$ is a relevant line tension and $\sigma$ is the surface tension, range from nanometers to micron. It is found that nano-scale texture causes a shift in the nucleation of ice in supercooled water to the three-phase contact line, while micro-scale texture does not.
\end{abstract}

\maketitle 

\paragraph*{}
While nucleation of solids in supercooled liquids is ubiquitous\cite{Diao2011,Sear2011,Sear2012}, surface nucleation, the tendency for freezing to begin preferentially at the liquid-gas interface, has remained puzzling\cite{Tab2002,Duft2004,ShawDurant2005,Shpyrko2006,Bowles2007,Sear2007,Sutter2007,
Djikaev2008,Carvalho2010}. Furthermore, in the presence of foreign catalysts the associated heterogeneous nucleation has been observed to prefer the three-phase contact line (triple line), especially for small particles \cite{Shaw2005} and rough surfaces \cite{Carvalho2010}.  Motivated by the conjectured importance of roughness and the contact line, we have searched for evidence of a shift to surface nucleation as the characteristic roughness length scale is decreased. Two plausible length scales associated with heterogeneous nucleation, the critical radius for a nucleation seed, and the length scale at which linear and surface energies are comparable, yield a range from micrometers to nanometers. In this Letter we show, using high speed imaging of the transient freezing process in supercooled water, that nano-scale texture causes a shift in the nucleation to the three-phase contact line, while micro-scale texture does not. Both the mean and variance of the freezing temperature are observed to increase, also pointing to the importance of nanotexture given that variances of independent causes add.  The possibility of a transition or optimal length scale has implications for the effectiveness of nucleation catalysts, including formation of ice in atmospheric clouds\cite{Cantrell2005}.

\paragraph*{}
Uniform probability of freezing is a standard assumption in nucleation theory: probability scaling as the volume of supercooled liquid for homogeneous nucleation, or as the area of the liquid--catalyst interface for heterogeneous nucleation. Recent studies suggest that for systems as widely varying as atomic liquids\cite{Bowles2007}, salts\cite{Bartell2000}, tetrahedral liquids\cite{Li2009}, hexaflourides\cite{Chushak2000}, metal alloys\cite{Shpyrko2006,Sutter2007}, Nickel-Silicon\cite{Lu2012}, polymers\cite{Carvalho2011}, and water\cite{Tab2002,Djikaev2002}, homogeneous nucleation prefers the liquid-vapor interface, and therefore its rate scales not as volume but rather as area. The mechanism for surface nucleation remains unclear, and even more troubling, its predominance has been qualified and questioned\cite{Duft2004,Sigurbjornsson2008,Knopf2006,Turner2005}. Meanwhile, experiments on the nucleation of ice on small particles in supercooled water have revealed a strong enhancement in nucleation rate for particles at the liquid-water -- air interface\cite{ShawDurant2005,Fornea2009}, suggesting that whatever physics underlies surface nucleation likely extends to heterogeneous nucleation as well. Sorting out this mystery is more than academic because it addresses fundamental aspects of classical nucleation theory (CNT) and thus predictability of nucleation processes; some long standing puzzles such as the empirical observation that `contact nucleation' is more efficient than `immersion nucleation' in supercooled cloud droplets\cite{endnote_definitions} may well be intertwined with the physics of surface nucleation. A leading hypothesis for the preference for surface nucleation is the formation of a three-phase interface\cite{Sear2007,Djikaev2008}, and this aspect is investigated here for heterogeneous nucleation of ice in supercooled water.

\paragraph*{}
Rough \cite{Carvalho2010} or `point-like contact' \cite{Shaw2005} nucleation catalysts have been observed to induce nucleation at the three-phase contact (triple) line. It has been suggested that a free energy per unit length or line tension $\tau$ for the contact line contributes to the nucleation kinetics \cite{DjikievRuckenstein2008}. Thus, an extensive nucleation rate (number of freezing events per unit time) would be a sum of contributions from immersion and contact modes. In recent work we sought direct confirmation by observing the freezing of mm-sized supercooled water droplets on atomically smooth substrates using high speed optical imaging: and yet for a variety of contact angles and cooling rates, no preference for nucleation at the macroscopic air-water-substrate contact line was observed \cite{Gurganus2011,Gurganus2013}. It is possible, however, that the lack of contact-line-nucleation in those experiments reflects the system geometry. For example, an extensive nucleation rate dependent on both droplet-substrate surface area and perimeter would lead to the relative role of immersion versus contact line nucleation scaling with drop diameter. If so, then decreasing the drop size should favor surface nucleation. Rather than decreasing the size of the supercooled liquid volume, which renders our high-speed imaging method more difficult, here we modify the geometry of the nucleation catalyst so as to impose `textures' exhibiting a range of length scales on the air-water-substrate contact line. The question is whether catalyst geometry alone can induce a preference for nucleation at the contact line.

\paragraph*{}
The apparent role of substrate geometry and roughness \cite{Carvalho2010,Diao2011} motivates a consideration of possible length scales that could enter the heterogeneous nucleation problem. One is the size of the critical nucleus predicted by CNT.  It has been shown \cite{Pound1964,Sear2006,Sear2010} that steps, pores, cracks, or other surface features with sizes on the order of the critical nucleus may promote more efficient nucleation by lowering the free energy barrier. For example, Page et al.\cite{Sear2006} demonstrated that a two-step nucleation rate exists for ice within and outside of a pore, and therefore an optimal pore size exists, near the critical nucleus size, at which nucleation rate is maximized. Quite generally, the critical radius for nucleation is obtained from the Gibbs-Thomson equation $r^\star = 2 \sigma v_i / \Delta \mu$, where $v_i$ is the molecular volume for ice and $\Delta \mu$ is the chemical potential difference between the supercooled liquid and the nucleated solid. It can be expressed as $\Delta \mu = kT \ln p_w/p_i \approx l_f \Delta T / T_0$, where $p_w$ and $p_i$ are the equilibrium vapor pressures of liquid water and ice, respectively, $l_f$ is the latent heat of fusion, $T_0$ is the melting temperature, and $\Delta T \equiv T_0 - T$ is the supercooling temperature. For the typical $\Delta T$ of 5 to 35 K, the critical radius varies over the approximate range $r^\star \approx 1$ to 10 nm. In the experiments reported here, the observed supercooling temperatures suggest a length scale $\lambda \approx 2 r^\star \approx 10$ nm as a candidate for substrate texture.

\begin{figure}[t!]
    \includegraphics[width=8.5cm]{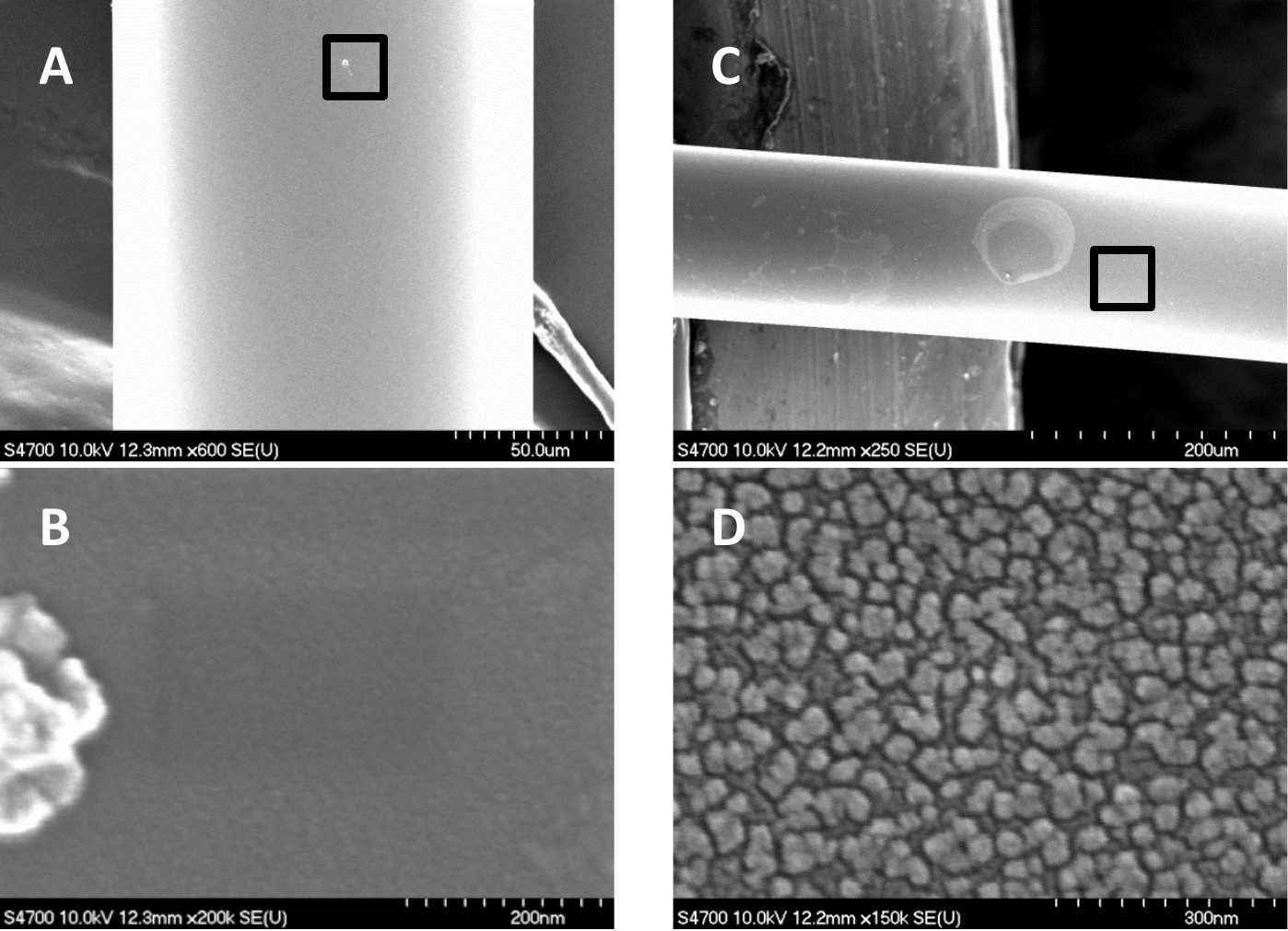}
    \caption{\textbf{Fabricating nanoscale surface texture.}  Motivated by the conjectured importance of roughness to heterogeneous nucleation and the plausible range of length scales, these experiments were conducted with smooth optical fibers (A,B), nano-textured optical fibers (C,D), and micro-textured silicon substrates (see supplemental Figure S2) as heterogeneous nucleation catalysts. Panels A-D were taken with a high resolution SEM.
}
      \label{Roughness}
\end{figure}

\paragraph*{}
A second length scale motivated by the suggested importance of the three-phase contact line, arises naturally from the notion that the contact line is characterized by a free energy per unit length, the line tension $\tau$. For a system involving air, supercooled liquid, nucleated solid, and catalyst substrate, four distinct line tensions exist and may play a role  \cite{Nav1981,DjikievRuckenstein2008}. Regardless of which $\tau$ or combination of $\tau$'s plays a role, the ratio of line and surface tension $\delta \sim \tau/\sigma$ suggests a length scale, below which free energy of the contact line exceeds free energy of the interface\cite{DeGennes}. The existence or significance of the line tension itself is still a matter of some debate\cite{Lohse2011,Rowlinson1982}, with conflicting reports in literature for the magnitude and even the sign\cite{Marmur1997,Berg2010}. Despite poor experimental quantification, however, recent computational\cite{Auer2003} and nucleation studies\cite{Kulmala2007} have reconciled observations to theory by including line tension. We ask, therefore, what substrate `texture' length scales would introduce geometric variability  to the contact line e.g., distortion due to pinning\cite{DeGennes}, that may affect the nucleation rate?  Perhaps surface texture length scales on the order of or smaller than $\delta$ will promote surface nucleation?  While $\sigma$ varies only slightly\cite{DeGennes} with $T$, from $10^{-2}-10^{-1}$  J m$^{-2}$, the reported range of values for $\tau$ is wide, from $10^{-11}$ to $10^{-8}$ J m$^{-1}$ \cite{endnote1}, yielding a range of $\delta$ from $<1$ nm to $\sim 1$ $\mu$m.

\paragraph*{}
In summary, length scales that could contribute to contact-line nucleation range from from the mm-scale of the macroscopic droplets for which no contact-line nucleation was observed, through the plausible range of $\delta$ starting at $\sim 1$ $\mu$m, and finally down to length scales of minimum $\delta$ and the critical nucleus size $1-10$ nm. To that end, we have conducted nucleation experiments in which we observe freezing of water with high-speed optical imaging to identify the spatial origin of nucleation with droplets in contact with surfaces that are textured over a range of length scales. A shift to preference for nucleation at the contact line in these experiments would suggest that, whatever the physical mechanism, catalyst geometry plays a defining role.  It then opens the way for further investigation of specific mechanisms using other methods, e.g., whether steps or pores resonant with the critical nucleus size, or distortion and curvature of the contact line on the order of $\delta$ lead to reduced Gibbs free energy barrier for nucleation.
\paragraph*{}
Guided by the cascade of scales described above, these experiments were conducted with heterogeneous nucleation catalysts textured to exhibit specific length scales. The fabricated catalysts consist of micro-textured silicon substrates (see supplemental Figure S2), and nano-textured optical fibers (Figure 1, panels C and D). Untextured substrates and fibers were used as controls; a smooth fiber is shown in Figure 1 (panels A and B) for reference. To impose micrometer-scales, single-crystal silicon substrates with periodic patterns of linear surface features were fabricated using photolithography (see supplemental material). The spatial feature sizes explored were 50, 10, 5, and 2 $\mu$m; For large etch depths a superimposed, random texture with lengths in the range 1 $\mu$m to $\sim 100$ nm also appeared (Supplemental Figure S2, bottom panel). To explore the nano-scales, below the limits of the photolithography method, an etching method was used on silica glass fibers. Fibers without and with the resulting nanotexture are shown at two resolutions in scanning electron microscope (SEM) images (Figure 1 A-D). Image analysis of the texture shown in Figure 1 (D) reveals linear sizes from approximately 100 nm down to 2 nm, which is near the resolution limit of the imaging method (see supplemental materials, section 4). Even in the absence of a contact line effect that changes with texture length scale, we can expect that roughness leads to an increase in catalyst surface area and therefore an increase in the extensive nucleation rate. The increases in surface area are small, but more importantly, the measurement depends on the spatial distribution of nucleation events, and is therefore is not directly dependent on quantification of nucleation rate.

\paragraph*{}
The freezing of supercooled water droplets in contact with a catalyst is observed with a high speed camera at 200 $\mu$s between frames.  The droplet is cooled at a rate of 2K min$^{-1}$ to a temperature below the droplet freezing temperature ($T_{Freeze}$), then warmed to 10 $^\circ$C to melt the droplet, see supplemental information for more details.  As shown in Figure 2, reversing the freezing process in time pinpoints the epicenter of crystallization. The process is repeated many times so that the spatial distribution of nucleation events can be measured. In each cycle the water droplet is cooled until freezing occurs and then heated and melted. The droplet rests on a substrate, as shown schematically in Figure 2. Looking from above, it is apparent whether there is a preference for nucleation events at the clearly visible three-phase contact line.  For smooth substrates it has been confirmed\cite{Gurganus2011,Gurganus2013} that nucleation events are distributed randomly with no spatial correlations or preference for the contact line. When glass fibers are examined, the fiber pierces the drop as shown in Figure 2. Examples of nucleation events initiated on the substrate (red), on the immersed fiber (green), and at the fiber contact line (blue) are illustrated. Because the substrate and the fiber have essentially the same chemical composition (silica), a spatial shift from the substrate to the immersed fiber or to the fiber contact line is considered evidence for a change in the nucleation efficiency of those regions that represent a negligibly small fraction of the total catalyst surface area.

\begin{figure}[t!]
    \includegraphics[width=7cm]{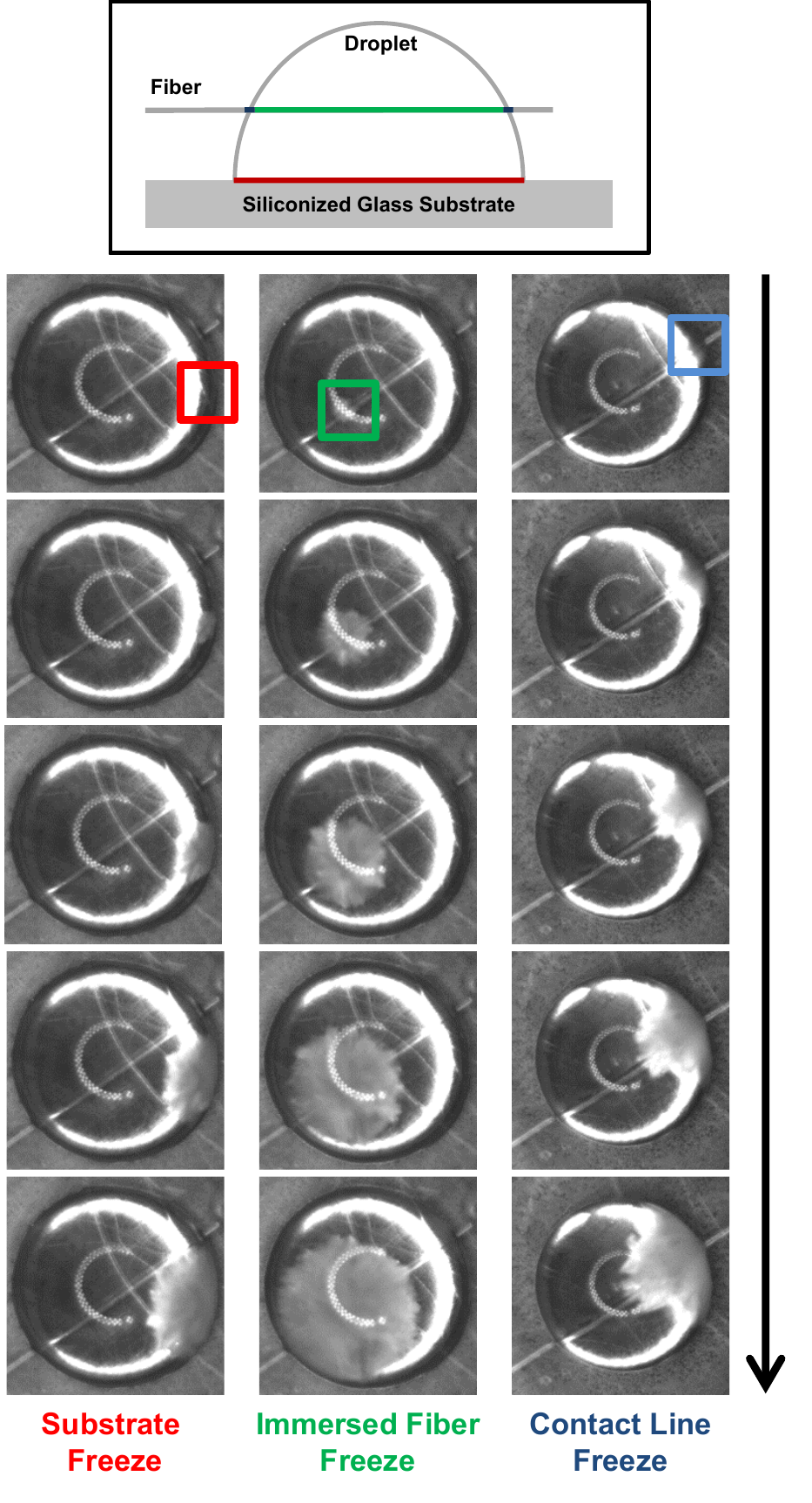}
    \caption{\textbf{Three modes of nucleation.}
    Top: A schematic of the droplet-fiber geometry.  A 30 $\mu$L droplet with a contact angle of $\approx 90^\circ$ rests on a siliconized glass slide (Hamilton Scientific) that is cooled from below \cite{Gurganus2013}.  An optical fiber, partially immersed within the droplet, can act as a heterogeneous nucleation catalyst.  Three possibilities for nucleation then arise: on the substrate (red), on the immersed fiber (green) and at the fiber contact lines (blue).  Bottom:  By imaging the crystallization at 5kHz we pinpoint the nucleation site (boxed area in film strips).  Film strips here represent each of the three nucleation modes. Every 15th frame is shown resulting in a 3 ms spacing.}
    \label{Three modes of nucleation}
\end{figure}

\begin{figure*}[]
    \includegraphics[width=18cm]{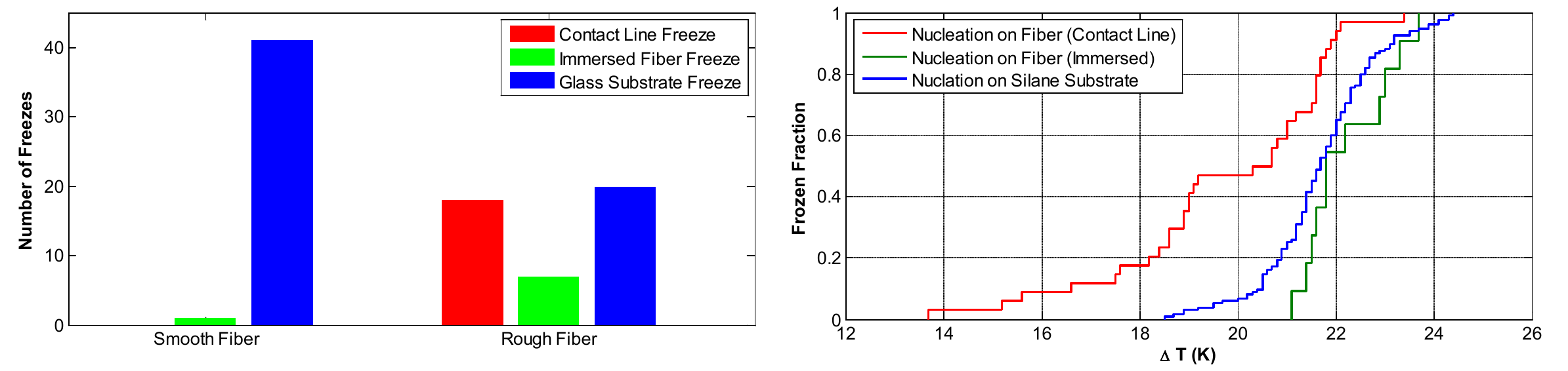}
    \caption{\textbf{Nano-Texture observed to cause a transition to surface nucleation at the contact line.}  Results for the three modes of nucleation, substrate, immersed fiber, fiber contact line (see Figure 2).  \textbf{Left Panel}: Spatial origin of crystallization is observed often to shift to the contact line for nano-textured (rough) fibers, but neither the smooth fiber ($r=70$ $\mu$m) nor the micro-textured substrates (2-100$\mu$m) yield such a shift (see supplemental material, section 2).  Despite relatively small surface area of the nano-textured fiber, over half of freezing events are initiated there.  Furthermore, despite the overwhelmingly small spatial odds, the majority of the fiber induced events originate at the contact line. \textbf{Right Panel}:  Higher freezing temperatures (weaker supercooling) are observed for nucleation events at the fiber contact line as evidenced by the cumulative freezing probabilities (red curve).  Broadening of the distribution accompanies fiber contact line events, as expected because of additional variability in the geometry of the nano-textured contact lines.}
    \label{Data}
\end{figure*}
\paragraph*{}
The microfabricated catalyst substrates, with length scales down to 2 $\mu$m for the imposed pattern, and down to 100 nm for the deep etches, showed no change in the spatial distribution of nucleation events. Similarly, when the glass fiber with radius of 70 $\mu$m was present, there was no tendency for nucleation to prefer the fiber over the substrate (see Figure 3, left panel). The nano-textured fiber, however displayed a shift in the spatial distribution of nucleation events to the fiber contact line. Despite the relatively small surface area of the nano-textured fiber, over half of the freezing events initiated there. And strikingly, despite the overwhelmingly small spatial odds, the majority of the fiber induced events originated at the three-phase contact line.  This shift is consistent both with the surface texture length scale approaching the most likely value of $\delta \sim 10$ nm, and the length scale associated with the critical radius for nucleation $\lambda \sim 10$ nm.  Of course, the spatial evidence alone cannot be considered direct evidence for one length versus the other.  Regardless of motivation, the observation clearly supports the notion that nano-scale surface features strongly favor ice nucleation at three-phase contact lines.
\paragraph*{}
In addition to the direct spatial evidence inherent to the design of the experiment, the temperature at which crystallization is initiated is also recorded, providing information on the efficiency of nucleation. Figure 3 (right panel) shows cumulative freezing probabilities versus $\Delta T$. Nucleation events at the fiber contact line (red curve) show significantly higher freezing temperatures, and these weaker supercoolings are indicative of nucleation rates enhanced by many orders of magnitude\cite{ShawDurant2005}. The freezing temperature distribution for contact line events is not only shifted to higher temperatures, but is broadened. This is consistent with expectations for surface variability\cite{Levine2013} because variance adds for independent causes (e.g., randomness inherent to nucleation and randomness associated with surface texture). The temperature distributions thus support the spatial evidence for surface texture inducing the change in freezing behavior.
\paragraph*{}
Is the observed temperature shift ($\delta T \approx 3K$) consistent with the proposed mechanisms?  Previous observations of contact line nucleation of ice suggest similar shifts of $\delta T = 2-5K$\cite{Shaw2005,Fornea2009}.  It is customary in CNT to represent the efficiency of a heterogeneous catalyst through the contact angle $\theta_o$ of the nucleated phase, assumed to have the shape of a spherical cap: smaller contact angle implies more efficient catalyst.  By comparison, to achieve $\delta T = 1K$ a  $\Delta \theta_o = 2^o$ is required \cite{endnote2} \cite{Ervens2013}.  Although line tension values for ice and water are poorly known, direct measurements of other substances, via the modified Young's equation $\cos{\theta_r} =  \cos{\theta_o} - \tau / \sigma r$, show that $r\sim10-100nm$ droplets exhibit $\Delta \theta \sim 10^o$\cite{Berg2010,Pompe2000}.  These values are consistent with our observed $\delta T$.  Could nanopores  explain this phenomenon?   The nucleation barrier for $r_{pore}\approx r^*$ has been shown to be a half of that for a flat catalyst.  CNT for ice in water can realistically result in $\delta T \sim 1-10K$ for similar changes in $\Delta G$.  However, this ``pore-enhancement'' can explain the magnitude of our observed temperature shift, but how they would cause a spatial transition to freezing at the contact line is unclear.
\paragraph*{}
This evidence for a significant role of surface texture and characteristic length scales has wide implications: from catalyst design for drug synthesis, to improved parametrization of ice nucleation in clouds within weather and climate models.  The demonstrated improvement in nucleation efficiency for nano-texture substrates is qualitatively consistent with recent work indicating that nucleation is enhanced by the introduction of sharp corners compared to circular shapes in catalysts with nano-pores\cite{Diao2011,Sear2011}.  The relevant length scale may be the radius of curvature of a wetted surface feature, which is much smaller for `square' nano-pores and is therefore consistent with the line tension hypothesis.  The results from this work also help clarify why past work with `point-like' contact\cite{ShawDurant2005} showed a strong preference for surface nucleation: It is likely that the naturally occurring, irregular, micron-scale particles used there have surface features on the order of or smaller than the line tension scale, $r<\delta$. This leads naturally to the speculation that spatially localized regions that are thought to induce crystallization, known as `active sites', may be associated with surface features (steps, kink sites, defects) with characteristic length scales at or below $\delta$.

\paragraph*{}
This research was supported in part by an award from the Department of Energy (DOE) Office of Science Graduate Fellowship Program, the DOE Atmospheric System Research program through grant DE-SC0006949, and by the NSF grant AGS-111916.  The authors wish to thank Dr. P. Bergstrom and the Microfabrication Facility at MTU for assistance with substrate lithography, and also Dr. J. Drelich and the Applied Chemical and Morphological Analysis Laboratory at MTU for assistance with AFM and SEM measurements.


\end{document}